\documentclass[11pt,letterpaper]{article}
\usepackage[margin=1in]{geometry}
\usepackage[hyphens]{url}
\usepackage{graphicx}
\usepackage{natbib}
\usepackage{caption}
\usepackage{microtype}
\usepackage{booktabs}
\usepackage{amsmath}
\usepackage{array}
\usepackage{float}
\usepackage{placeins}
\title{Beyond Predictive Accuracy: A Reliability-Aware Audit of Molecular
Representations for Human Olfaction}
\author{
Kai Lun Huang\\
Department of Electrical and Computer Engineering\\
California State University, Fullerton\\
Fullerton, CA, USA\\
\texttt{kl.huang@csu.fullerton.edu}
\and
Wei Chieh Sun\\
Department of Electrical \& Computer Engineering\\
University of Washington\\
Seattle, WA, USA\\
\texttt{wsun12@uw.edu}
}
\date{}

\begin{document}
\maketitle

\begin{abstract}
Pretrained molecular encoders are commonly evaluated through downstream
prediction, but predictive accuracy alone does not establish that a learned
representation captures reproducible scientific structure, adds information
beyond strong conventional baselines, or transfers out of distribution. We
present a reliability-aware audit of generic molecular representations for
human olfaction across four distinct claims: global perceptual geometry,
incremental predictive value beyond chemistry, cross-dataset replication, and
mixture transfer to unseen components.

Using the Keller--Vosshall and Bierling single-molecule rating datasets and the
Ma binary-mixture dataset, we compare MoLFormer and ChemBERTa against RDKit
descriptors and Morgan fingerprints under identity-controlled and matched
evaluations. Human three-attribute rating geometry, based on intensity,
pleasantness, and familiarity, is reproducible across participant splits
(median RSA 0.743 and 0.855), whereas model--human alignment is substantially
weaker (RSA 0.019--0.158). Learned embeddings do not consistently outperform
conventional representations in global alignment, and MoLFormer provides no
clear incremental predictive value beyond a combined RDKit--Morgan baseline in
either single-molecule dataset. Human geometry shows positive but incomplete
agreement across 63 shared molecules (RSA 0.331; 95\% bootstrap interval
[0.204, 0.507]). Under one strict unseen-component mixture split, incremental
effects are outcome- and representation-dependent, with all intervals crossing
zero.

These results establish empirical boundaries for the evaluated generic
molecular encoders and motivate a broader evaluation principle: representation
quality in scientific domains should be assessed separately for target
reliability, structural alignment, incremental information, replication, and
out-of-distribution transfer.
\end{abstract}

\section{Introduction}
Scientific representations are often evaluated through downstream predictive
accuracy. That criterion is useful but incomplete: it does not establish that the
target itself is reproducible, that representation distances preserve target
structure, that learned features add information beyond complementary domain
baselines, that findings replicate across measurement protocols, or that they transfer
out of distribution. These claims require different controls and analysis units.

Human olfaction provides a demanding case study. Chemical similarity, perceptual
similarity, single-outcome prediction, and mixture behavior are not interchangeable.
Molecular features can predict pleasantness, descriptor profiles, and perceptual
similarity~\cite{khan_pleasantness_2007,snitz_similarity_2013,keller_dream_2017}, yet
human judgments vary across protocols and datasets
~\cite{keller_vosshall_2016,bierling_scientific_data_2025}. Mixtures introduce a
further integration problem that is not reducible to single-molecule prediction
~\cite{laing_francis_1989,bushdid_2014,ma_binary_mixtures_2021}. Generic molecular
encoders may preserve useful chemistry without reproducing human perceptual
organization, and apparent gains may reflect information omitted by an incomplete
baseline.

We ask: what aspects of human olfactory ratings are captured by generic pretrained
molecular representations, and do those signals extend beyond conventional chemical
features, replicate across datasets, and transfer to mixtures with unseen components?
We define the human target narrowly as \emph{the evaluated three-attribute perceptual
rating geometry based on intensity, pleasantness, and familiarity}. Global
representational similarity analysis (RSA) tests whether a representation preserves
the overall rank ordering of distances among all molecule pairs
~\cite{kriegeskorte_rsa_2008}. Incremental prediction tests whether an embedding adds
information after chemistry is already available. Cross-dataset comparison and strict
component-disjoint mixture evaluation test two different forms of transfer.
Predictive accuracy in one dataset does not establish reproducible perceptual geometry,
incremental information beyond conventional chemistry, robustness across measurement
protocols, or transfer to mixtures with unseen components.

We use the Keller--Vosshall and Bierling et al. single-molecule datasets
~\cite{keller_vosshall_2016,bierling_scientific_data_2025} and the Ma et al.
binary-mixture dataset~\cite{ma_binary_mixtures_2021}. MoLFormer and ChemBERTa are
compared with RDKit descriptors and Morgan fingerprints
~\cite{ross_molformer_2022,chithrananda_chemberta_2020,rdkit_software_2025,rogers_hahn_2010}.
We instantiate a general claim--evidence audit through five contributions:
\begin{itemize}
\item a reliability-aware audit that separates target reproducibility from
model--human global alignment;
\item incremental testing against complementary RDKit--Morgan domain baselines;
\item replication across independent single-molecule rating datasets;
\item strict component-disjoint transfer to mixtures containing unseen components;
and
\item reusable artifacts containing mappings, splits, configurations,
validated result tables, and figure-generation code.
\end{itemize}

\begin{figure*}[!t]
  \centering
  \includegraphics[width=\textwidth]{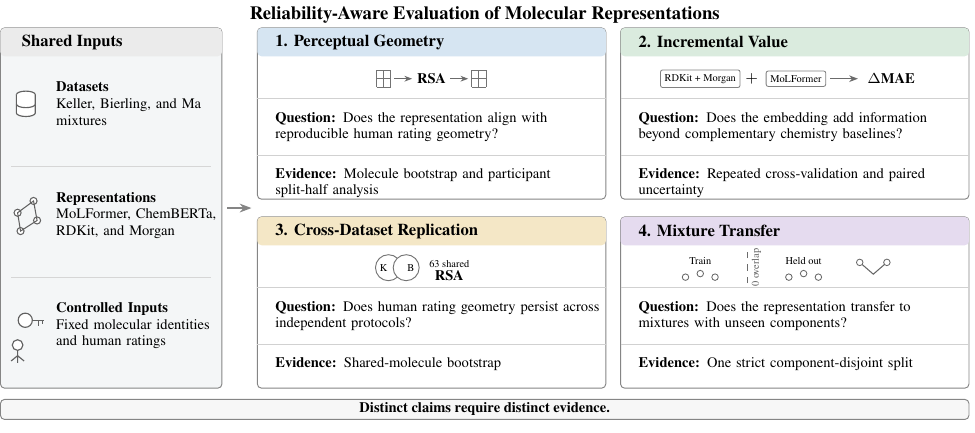}
  \caption{Reliability-aware evaluation of molecular representations for human
  olfaction. Shared datasets, molecular identities, representations, and human
  ratings support four distinct claims; evidence for one claim is not treated as
  validation of the others.}
  \label{fig:experimental_framework}
\end{figure*}

The resulting evaluation design is broader than this application: it makes explicit
which evidence supports reliability, structural alignment, incremental information,
replication, and out-of-distribution transfer. Figure~\ref{fig:experimental_framework}
shows its instantiation for human olfaction.

\section{Related Work}
Representation evaluation spans probing, benchmark construction, baseline design, and
out-of-distribution assessment. A downstream probe can reveal accessible information
without showing that an embedding preserves the global organization of a scientific
target; likewise, benchmark conclusions depend on controls that capture strong domain
knowledge and on splits that match the intended deployment shift. Scientific
representation learning therefore benefits from separating in-distribution prediction,
structural alignment, cross-protocol replication, and transfer. Our study applies this
evaluation logic to existing olfactory datasets and representations; it introduces
neither a new dataset nor a universal benchmark.

\subsection{Molecular Representations for Olfactory Prediction}
Olfactory prediction has a long connection to molecular features. Structure-derived
features have been used to predict odor pleasantness, perceptual similarity, and
descriptor ratings~\cite{khan_pleasantness_2007,snitz_similarity_2013,keller_dream_2017}.
These studies motivate molecular modeling of human smell, but they also show why
strong chemical controls are necessary: conventional descriptors can already capture
some odor-relevant structure. In this paper, RDKit descriptors and Morgan fingerprints
provide complementary two-dimensional chemistry blocks~\cite{rdkit_software_2025,rogers_hahn_2010}.
Their combination is the primary stringent baseline for incremental prediction;
the RDKit-only comparison is retained as a sensitivity analysis. Apparent gains from
learned embeddings are therefore interpreted only relative to these specified controls.

Learned molecular representations provide a broader feature family. Molecular graph
networks learn structure-aware features from atom-bond graphs
~\cite{duvenaud_molecular_fingerprints_2015,gilmer_mpnn_2017}, while molecular
benchmarks and unsupervised representations such as MoleculeNet and Mol2vec helped
standardize reusable molecular feature learning~\cite{wu_moleculenet_2018,jaeger_mol2vec_2018}.
SMILES-based language models adapt sequence-modeling ideas from transformer and BERT
architectures~\cite{weininger_smiles_1988,vaswani_attention_2017,devlin_bert_2019}.
MoLFormer and ChemBERTa are examples of generic pretrained molecular encoders trained
for molecular property prediction rather than olfaction-specific inference
~\cite{ross_molformer_2022,chithrananda_chemberta_2020}. Olfaction-specific work,
including machine-learning approaches to scent and the principal odor map, suggests
that learned representations can encode odor-relevant structure
~\cite{sanchez_lengeling_scent_2019,lee_principal_odor_map_2023}. The unresolved
question is whether public generic representations align with the evaluated perceptual
rating geometry, add information beyond conventional chemistry, and show agreement
across independent datasets. The present conclusions concern the evaluated generic
molecular encoders and do not determine the validity of olfaction-specific,
receptor-informed, graph-based, or three-dimensional representations.

\subsection{Perceptual Structure, Replication, and Mixture Generalization}
Outcome prediction and perceptual geometry are different targets. RSA compares
distance structures rather than single labels~\cite{kriegeskorte_rsa_2008}, and
olfactory work has used descriptor spaces and perceptual similarity to study the
organization of smell judgments~\cite{castro_descriptor_space_2013,koulakov_olfactory_space_2011}.
This distinction matters for representation evaluation: a model may predict intensity
without matching the evaluated intensity--pleasantness--familiarity geometry.

Replication is equally important because olfactory datasets differ in participants,
rating scales, molecule sets, and aggregation procedures. The Keller--Vosshall dataset
provides chemically diverse single molecules~\cite{keller_vosshall_2016}; the
Bierling et al. dataset provides an independent monomolecular dataset from a larger layperson sample
~\cite{bierling_scientific_data_2025,bierling_2025}. Comparing these datasets tests
whether representation effects are tied to one measurement protocol or survive an
independent human dataset.

Mixture perception poses a stricter compositional problem. Classic psychophysical work
showed that identifying odor components in mixtures is limited~\cite{laing_francis_1989},
and Bushdid et al. used complex mixtures to study human olfactory discrimination
capacity~\cite{bushdid_2014}. The Ma et al. binary-mixture dataset provides intensity
and pleasantness ratings for 222 mixtures of 72 food odorants~\cite{ma_binary_mixtures_2021}.
Random splits and pair-grouped splits can test interpolation among observed components
or pairs, but they do not establish generalization to mixtures containing unseen
molecular components. The strict unseen-component split used here is therefore this
paper's stringent mixture-transfer evaluation setting. Together, these datasets permit
controlled tests of perceptual rating geometry, incremental value beyond chemistry,
cross-dataset agreement, and transfer under a strict unseen-component split.

\section{Data and Evaluation Design}
\subsection{Datasets, Cohorts, and Molecular Identity}
Keller--Vosshall contributes 55 eligible participants and a final eligible set of
476 molecules, with intensity as the prediction target and intensity, pleasantness,
and familiarity defining the three-attribute geometry~\cite{keller_vosshall_2016}.
Bierling contributes 73 stereo-aware molecules
~\cite{bierling_scientific_data_2025,bierling_2025}. Its primary cohort follows the
source dictionary and notebook: main-study records with \texttt{inclusion=1}, excluding
the patient sampling group and retest rows. This yields 1,119 participants and 11,190
rating rows. The source odor metadata list 74 odor codes, but \texttt{4Isoprop}
(cuminol; CID 325) has no rating row; the other 73 have resolved identities and enter
the molecule-level analysis. The previous broader cohort is retained only as a
cohort-definition sensitivity, and model performance did not determine cohort choice.
Ma contributes 72 components, 30 trained assessors, and 222 mixture units aggregated
to $I_{AB}$ and $P_{AB}$ outcomes~\cite{ma_binary_mixtures_2021}.

\begin{table*}[t]
\centering
\caption{Datasets and primary analysis units.}
\label{tab:datasets}
\footnotesize
\setlength{\tabcolsep}{4pt}
\newcommand{\tblcell}[2]{\parbox[t]{#1}{\raggedright #2}}
\begin{tabular}{@{}llllll@{}}
\toprule
\tblcell{.07\textwidth}{Dataset} & \tblcell{.16\textwidth}{Scientific role} & \tblcell{.12\textwidth}{Molecules / components} & \tblcell{.14\textwidth}{Participants / assessors} & \tblcell{.11\textwidth}{Analysis units} & \tblcell{.31\textwidth}{Outcomes} \\
\midrule
\tblcell{.07\textwidth}{Keller} & \tblcell{.16\textwidth}{Single-molecule evaluation} & \tblcell{.12\textwidth}{476 molecules} & \tblcell{.14\textwidth}{55 participants} & \tblcell{.11\textwidth}{476 molecules} & \tblcell{.31\textwidth}{Intensity prediction; intensity, pleasantness, and familiarity geometry} \\
\tblcell{.07\textwidth}{Bierling} & \tblcell{.16\textwidth}{Independent replication} & \tblcell{.12\textwidth}{73 molecules} & \tblcell{.14\textwidth}{1,119 participants} & \tblcell{.11\textwidth}{73 molecules} & \tblcell{.31\textwidth}{Intensity, pleasantness, and familiarity} \\
\tblcell{.07\textwidth}{Ma} & \tblcell{.16\textwidth}{Strict mixture transfer} & \tblcell{.12\textwidth}{72 components} & \tblcell{.14\textwidth}{30 assessors} & \tblcell{.11\textwidth}{222 mixtures} & \tblcell{.31\textwidth}{$I_{AB}$ intensity; $P_{AB}$ pleasantness} \\
\bottomrule
\end{tabular}
\let\tblcell\relax
\end{table*}

Fixed stereo-aware molecular identity mappings prevent duplicate molecules,
cross-dataset mismatch, and component leakage. Representations and outcomes were
joined by molecular identity rather than row position.

\subsection{Representations and Perceptual Geometry}
RDKit 2025.09.2 generated 217 two-dimensional descriptors. Geometry used median-filled,
z-scored descriptors and cosine distance; prediction fitted mean imputation and
standardization within each training fold. Morgan fingerprints were 2,048-bit vectors
(radius 2, chirality enabled) generated with the same RDKit version; Tanimoto distance
was one minus bit-vector Tanimoto similarity, and prediction left binary bits unscaled
~\cite{rdkit_software_2025,rogers_hahn_2010}.

Frozen MoLFormer-XL-both-10pct embeddings used the model's
768-dimensional pooled output, while ChemBERTa-77M-MLM embeddings used
attention-mask-weighted mean pooling over the final hidden states to obtain
384 dimensions. Both representations were precomputed once from canonical
SMILES under deterministic evaluation settings, and neither encoder was
fine-tuned. Exact checkpoint revisions, tokenizer settings, and extraction
configurations are provided in the Supplementary Document and Code and Data
Supplement.
MoLFormer is the learned representation used for incremental prediction, while
ChemBERTa provides a second generic encoder in geometry and mixture comparisons, as
documented in the representation registry.

For each dataset, molecule-level intensity, pleasantness, and familiarity were
z-scored across molecules; Euclidean distance defined the human rating matrix. Learned
embeddings used cosine distance. RSA was the Spearman correlation between strict upper
triangles:
\begin{equation}
\begin{aligned}
\rho_{\mathrm{RSA}}=\operatorname{Spearman}\!\big(&
\operatorname{vec}_{\triangle}(\mathbf D_{\mathrm{rep}}),\\
&\operatorname{vec}_{\triangle}(\mathbf D_{\mathrm{human}})\big).
\end{aligned}
\label{eq:rsa}
\end{equation}
For molecule bootstrap intervals, molecules were sampled with replacement, both
distance matrices were reconstructed for each sample, and RSA was recalculated from
their strict upper triangles. We used 2,000 valid replicates and percentile intervals;
molecule pairs were not treated as independent. In the retained geometry-null control,
representation-matrix molecule labels were permuted 2,000 times while the human matrix
was held fixed, and RSA was recomputed after each permutation. This control is distinct
from predictive evaluation.

\subsection{Empirical Participant Split-Half Reproducibility}
Participants were assigned to independent halves, and molecule means were reconstructed
separately on the same molecule set. Each half produced a three-attribute Euclidean
geometry whose upper triangles were compared by Spearman RSA. Bierling splits were
stratified by source-design variables; the primary cohort contains no retest rows.
Using a fixed master seed, we obtained 1,000 valid splits per dataset. This estimates
empirical reproducibility of the aggregate geometry, not a theoretical ceiling.

\subsection{Incremental Predictive Validity}
The stringent comparison is RDKit + Morgan versus RDKit + Morgan + MoLFormer. All
feature sets used the same molecule-level folds: five folds, ten repeats, and master
seed 20260713. Every molecule therefore received ten out-of-fold predictions; these
were averaged per molecule before MAE was calculated. Training-fold preprocessing used
mean imputation for all blocks, standardization for RDKit and MoLFormer, and no scaling
for Morgan bits. Blocks were concatenated without weighting and shared one Ridge
penalty. Ridge $\alpha=10$ was fixed under the existing analysis policy rather than
selected from the full data. The nonlinear sensitivity used 80-tree Random Forests
(maximum depth 10, minimum leaf size 2, \texttt{max\_features=sqrt}, seed 20260713)
~\cite{breiman_random_forests_2001}.

Paired uncertainty used 2,000 molecule bootstrap resamples of the averaged out-of-fold
predictions and 2.5th--97.5th percentile intervals. We define
\begin{equation}
\begin{aligned}
\Delta\mathrm{MAE}={}&\mathrm{MAE}(\mathrm{Chemistry+MoLFormer})\\
&-\mathrm{MAE}(\mathrm{Chemistry}),
\end{aligned}
\label{eq:delta_mae}
\end{equation}
so negative values are beneficial. RDKit-only comparisons are secondary.

\subsection{Cross-Dataset and Mixture Evaluation}
Cross-dataset geometry used 63 exact shared stereo-aware molecules in identical order.
Each dataset was z-scored separately. The shared-molecule bootstrap resampled molecules,
reconstructed both geometries, and used 2,000 percentile-bootstrap replicates rather
than treating the 1,953 pairwise distances as independent.

For mixtures, component vectors were mean-pooled under the primary rule. The
prespecified strict split has 52 training components, 11 held-out components, zero
overlap, 101 training units, 20 test units, and 101 excluded cross-partition units.
Mixtures joining training- and test-side components were excluded. A fixed,
outcome-blind search of 5,000 assignments assessed availability of another partition
with exactly 11 held-out components, at least 20 strict test units, and at least 101
strict training units. None qualified, so no repeated-partition model was fit.

\section{Results}
\subsection{Human Split-Half Reproducibility and Model--Human Alignment}
\begin{figure*}[!t]
  \centering
  \includegraphics[width=\textwidth]{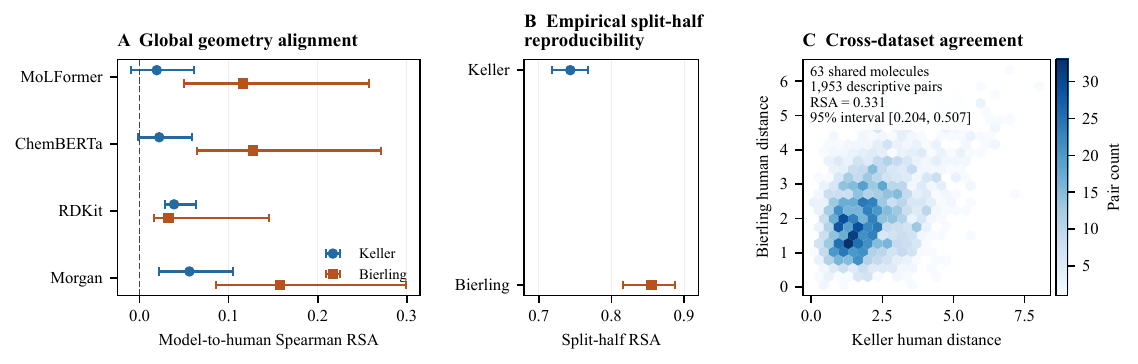}
  \caption{Geometry results. (A) Model-to-human global RSA; whiskers are 95\%
  molecule-bootstrap intervals, including intervals from 2,000 primary-cohort
  Bierling resamples. (B) Empirical split-half reproducibility; whiskers are the
  2.5th--97.5th percentiles across participant splits and are not model-performance
  ceilings. (C) Descriptive density of Keller and Bierling human distances across 63
  shared molecules. Its interval resamples shared molecules, not the 1,953 dependent
  pairwise distances. These three uncertainty summaries arise from different
  resampling procedures and are not interchangeable.}
  \label{fig:geometry_alignment}
\end{figure*}

The evaluated three-attribute geometry was substantially reproducible across
participant splits (Figure~\ref{fig:geometry_alignment}B). Across 1,000 valid splits,
median RSA was 0.743 in Keller (2.5th--97.5th split percentiles 0.719--0.768) and
0.855 in Bierling (0.816--0.888). These are empirical participant split-half
reproducibility estimates, not theoretical ceilings.

Model-to-human alignment was much weaker (Figure~\ref{fig:geometry_alignment}A). In
Keller, RSA was 0.019 for MoLFormer, 0.022 for ChemBERTa, 0.039 for RDKit, and 0.056
for Morgan. In Bierling, RSA and 95\% molecule-bootstrap intervals were 0.116
[0.050, 0.258], 0.127 [0.065, 0.272], 0.033 [0.016, 0.146], and 0.158
[0.086, 0.299], respectively. Morgan had the highest Bierling point estimate,
although the molecule-bootstrap intervals overlapped substantially across
representations. Learned embeddings therefore did not consistently dominate
conventional chemistry.

Global RSA evaluates preservation of the overall rank ordering of all molecule-pair
distances. Low RSA indicates weak global agreement under the specified distances, not
complete absence of outcome-specific information. The separation between empirical
split-half RSA and model RSA also makes participant-level unreliability in aggregate
ratings an unlikely primary explanation for weak model alignment.

\subsection{Sensitivity to Cohort Definition}
The primary Bierling cohort was determined from the source study design rather than
model performance. The previous broader cohort contained 1,314 participant IDs and
13,260 rows; the primary main-study, included, non-patient cohort without retest rows
contains 1,119 participants and 11,190 rows. Both retain 73 molecules. Molecule-level
outcomes were highly concordant (Spearman 0.981 for intensity, 0.994 for
pleasantness, and 0.979 for familiarity), and the two perceptual distance matrices had
RSA 0.963 (Table~\ref{tab:cohort_sensitivity}). Representation RSA changed modestly,
and the qualitative geometry conclusion was unchanged.

\begin{table*}[t]
\centering
\caption{Cohort-definition sensitivity. Agreement columns compare molecule-level
outcomes or distance matrices; model rows report RSA within each cohort.}
\label{tab:cohort_sensitivity}
\small
\resizebox{\textwidth}{!}{%
\begin{tabular}{lrrl@{\qquad}lrrl}
\toprule
Quantity & Broader & Primary & Agreement/change & Quantity & Broader & Primary & Change \\
\midrule
Participants & 1,314 & 1,119 & $-195$ & Split-half median & 0.867 & 0.855 & $-0.012$ \\
Rating rows & 13,260 & 11,190 & $-$2,070 & MoLFormer RSA & 0.111 & 0.116 & $+0.005$ \\
Molecules & 73 & 73 & same set & ChemBERTa RSA & 0.123 & 0.127 & $+0.004$ \\
Intensity outcome & -- & -- & $\rho=0.981$ & RDKit RSA & 0.029 & 0.033 & $+0.003$ \\
Pleasantness outcome & -- & -- & $\rho=0.994$ & Morgan RSA & 0.151 & 0.158 & $+0.007$ \\
Familiarity outcome & -- & -- & $\rho=0.979$ & Shared-geometry RSA & 0.352 & 0.331 & $-0.021$ \\
Distance matrix & -- & -- & RSA $=0.963$ & & & & \\
\bottomrule
\end{tabular}%
}
\end{table*}

\subsection{Partial Cross-Dataset Agreement in Human Rating Geometry}
Across 63 shared stereo-aware molecules, Keller and primary-cohort Bierling
three-attribute distance matrices showed positive but incomplete agreement
(Spearman RSA $=0.331$, 95\% shared-molecule bootstrap interval [0.204, 0.507];
Figure~\ref{fig:geometry_alignment}C). The bootstrap resampled shared molecules and
reconstructed both geometries; the 1,953 plotted pairwise distances are descriptive,
not independent analysis units.

The datasets differed quantitatively in representation alignment and prediction.
Molecule composition, participant design, protocol, and outcome aggregation may all
contribute. The shared-molecule result is protocol-level agreement between aggregate
geometries; it is neither participant reliability nor a model-performance ceiling.

\subsection{No Clear Increment Beyond Conventional Chemistry}
\begin{figure*}[!t]
  \centering
  \includegraphics[width=\textwidth]{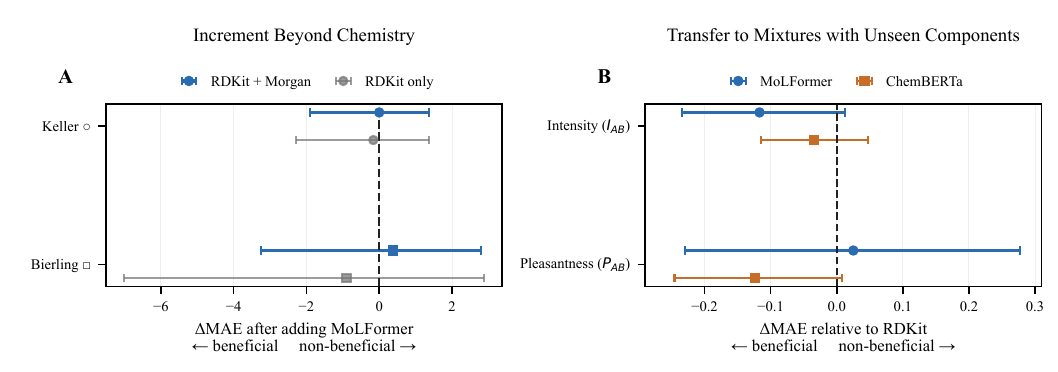}
  \caption{Predictive value under increasingly stringent controls. (A) MoLFormer
  increment beyond RDKit--Morgan and RDKit-only chemistry baselines in repeated
  single-molecule evaluation; whiskers are paired molecule-bootstrap intervals.
  (B) Incremental effects relative to RDKit under one prespecified strict
  unseen-component mixture split; whiskers are mixture-unit bootstrap intervals.
  Negative $\Delta$MAE is beneficial. The panels provide distinct forms of evidence.}
  \label{fig:predictive_transfer}
\end{figure*}

The stringent incremental comparison asks whether MoLFormer improves prediction after
both RDKit descriptors and Morgan fingerprints are available
(Figure~\ref{fig:predictive_transfer}A). In Keller, MAE changed from 13.369 to 13.368,
giving $\Delta$MAE $=-0.001$ (95\% molecule-bootstrap interval $-1.911$ to 1.365).
The point estimate was effectively zero. In Bierling, MAE changed from
9.734 to 10.111, giving $\Delta$MAE $=+0.377$ (interval $-3.240$ to 2.784), a
non-beneficial point estimate. Both intervals were broad and crossed zero.

The Random Forest sensitivity was also non-beneficial: $\Delta$MAE was +0.200 in
Keller and +0.576 in Bierling. We therefore find no clear incremental predictive value
from MoLFormer beyond the combined RDKit--Morgan baseline in either dataset.

Against RDKit alone, retained as a secondary diagnostic, Ridge point estimates were
directionally beneficial but uncertain: $-0.163$ in Keller (interval $-2.282$ to
1.366) and $-0.902$ in Bierling (interval $-7.017$ to 2.885). These secondary values
do not establish improvement and are not the paper's primary incremental claim.

\subsection{Outcome-Dependent Mixture Transfer Under One Strict Split}
The prespecified strict unseen-component split contains 52 training components, 11
held-out components, zero component overlap, 101 training units, 20 test units, and
101 excluded cross-partition units. All conclusions here are limited to this one split
(Figure~\ref{fig:predictive_transfer}B; Table~\ref{tab:mixture_exact}).

For intensity $I_{AB}$, RDKit MAE was 0.4395. MoLFormer alone achieved 0.3157 and
RDKit + MoLFormer 0.3226, giving $\Delta$MAE $=-0.1169$ (95\% mixture-unit bootstrap
interval $-0.2347$ to 0.0118). ChemBERTa alone achieved 0.3409 and RDKit + ChemBERTa
0.4050; its $\Delta$MAE was $-0.0344$ ($-0.1139$ to 0.0478). For pleasantness
$P_{AB}$, RDKit MAE was 0.5998. MoLFormer alone achieved 0.5710 and the combined model
0.6248, a non-beneficial $\Delta$MAE of +0.0251 ($-0.2293$ to 0.2769). ChemBERTa
alone achieved 0.6518 and the combined model 0.4761; its $\Delta$MAE was $-0.1237$
($-0.2455$ to 0.0085). Every interval crossed zero.

Thus direction differed by outcome and learned representation. The primary MoLFormer
comparison was directionally beneficial for intensity but not pleasantness; the
secondary ChemBERTa comparison showed the reverse ordering in magnitude. These point
estimates do not establish a representation or outcome difference.

\begin{table*}[!t]
\centering
\caption{Strict-split mixture results (20 test units). Intervals use the validated
mixture-unit bootstrap; negative $\Delta$MAE is beneficial.}
\label{tab:mixture_exact}
\small
\begin{tabular}{@{}llrrl@{}}
\toprule
Outcome & Added & RDKit & Combined & $\Delta$MAE [95\% CI] \\
\midrule
$I_{AB}$ & MoLFormer & 0.4395 & 0.3226 & $-0.1169$ [$-0.2347$, 0.0118] \\
$I_{AB}$ & ChemBERTa & 0.4395 & 0.4050 & $-0.0344$ [$-0.1139$, 0.0478] \\
$P_{AB}$ & MoLFormer & 0.5998 & 0.6248 & $0.0251$ [$-0.2293$, 0.2769] \\
$P_{AB}$ & ChemBERTa & 0.5998 & 0.4761 & $-0.1237$ [$-0.2455$, 0.0085] \\
\bottomrule
\end{tabular}
\end{table*}

An outcome-blind search over 5,000 fixed assignments found no additional partition
matching the explicit criteria. Of 4,936 unique test-component sets, 66 had exactly 11
test components, but none reached 20 strict test units (maximum 14). No
repeated-partition model was fit. This limits robustness assessment without showing
that no matching partition exists in the complete combinatorial graph.

\section{Discussion}
\subsection{Empirical Boundaries of Generic Molecular Encoders}
This study separates global perceptual geometry, outcome-specific prediction,
increment beyond chemistry, independent-dataset agreement, and compositional transfer.
Aggregate three-attribute human rating geometry was substantially reproducible, while
model-to-human global alignment was much weaker. Learned embeddings did not
consistently dominate conventional chemistry, and MoLFormer provided no clear
incremental value beyond RDKit + Morgan.

Cross-dataset human geometry showed positive but incomplete agreement, with a broad
shared-molecule interval. Differences in molecule composition, participant design,
protocol, and aggregation may contribute; these comparisons do not identify causes.
Under one strict split, mixture effects depended on outcome and representation. Thus,
generic molecular encoders may preserve useful chemical structure without
reconstructing the evaluated human organization or adding information beyond strong
conventional baselines.

\subsection{Implications for Molecular Representation Learning}
At minimum, scientific representation evaluation should (i) estimate reproducibility
of the target, (ii) test increment beyond complementary domain baselines, and (iii)
separate in-dataset prediction, global geometry, independent-protocol replication, and
out-of-distribution transfer. Applying these controls prevents evidence for one claim
from being treated as evidence for all of them.

\paragraph{Global geometry and outcome-specific prediction.}
Global RSA asks whether all pairwise perceptual-distance ranks are preserved. A useful
predictive direction for one outcome need not preserve that full ordering. Weak
geometry therefore does not imply complete absence of useful information, but
predictive accuracy in one setting does not validate geometry, chemical increment,
cross-protocol robustness, or mixture transfer.

\paragraph{Baseline completeness.}
MoLFormer's directionally beneficial but uncertain RDKit-only point estimates
disappeared under RDKit + Morgan: Keller was essentially zero and Bierling was
non-beneficial. This is consistent with substantial redundancy between generic learned
embeddings and conventional two-dimensional chemistry. Claims of perceptual increment
therefore depend strongly on baseline completeness. RDKit + Morgan is a stringent
empirical control here, not an information-theoretic ceiling. The observed pattern does
not establish representational equivalence. Olfaction-specific,
receptor-informed, graph-based, and three-dimensional representations remain outside
the evaluated scope.

\paragraph{Mixture outcomes.}
The strict-split mixture pattern depended on outcome and representation. Intensity may
be more compatible with simple component aggregation, whereas pleasantness may depend
more strongly on configural or interaction effects. This interpretation is tentative:
one small split cannot establish a mechanism, and the absence of a second structurally
matched partition prevents a robustness claim. Future mixture studies require larger
component-disjoint evaluations, concentration information, and interaction-aware
composition models.

\section{Limitations and Ethical Considerations}
This study uses public secondary human-subject datasets and recruits no participants.
It inherits source consent, sampling, population, and governance limitations; no new
IRB status is claimed. Rating scales, dilution handling, protocols, participant design,
outcome aggregation, and molecule composition differ across datasets and limit causal
or direct cross-dataset interpretation.

The evaluated human geometry contains only intensity, pleasantness, and familiarity
and does not define complete olfactory perception. Participant split-half reliability
is empirical, not a theoretical ceiling. The primary Bierling main, included,
non-patient cohort differs from the previous broader cohort, although the cohort audit
found qualitatively similar conclusions. Only generic molecular encoders were evaluated;
no confirmed olfaction-specific pretrained representation was included, and conclusions
must not generalize to all molecular representations.

The mixture evidence comes from one small component-disjoint test partition with 20
units, which may remain dependent through components shared within the test set. No
additional partition in the fixed 5,000-candidate search matched its 11 held-out
components, 20-test-unit, and 101-training-unit criteria. Concentration weights were unavailable, and mixture conclusions are
sensitive to outcome, composition rule, baseline, and probe. These limitations preclude
generalization to all mixtures or all aspects of olfactory perception.

\paragraph{Code availability.}
Code, configurations, molecular mappings, split assignments, and derived
verification artifacts will be released publicly following completion of the
review and archival process. Raw source datasets and pretrained model weights
are not redistributed and remain available from their original providers.

\paragraph{Use of Generative AI.} Generative AI tools were used in an assistive role for code drafting and
debugging, documentation organization, and language editing. All generated code
and text were reviewed, all analyses were executed on the reported data and
codebase, and all methodological decisions, numerical results, and scientific
interpretations were verified by the authors, who retain full responsibility
for the work.

\section{Conclusion}
Aggregate three-attribute human rating geometry was substantially reproducible, but the
generic molecular representations evaluated here showed weak global alignment and did
not consistently outperform conventional chemistry. MoLFormer provided no clear
incremental predictive value beyond the combined RDKit--Morgan baseline. Under one
prespecified strict unseen-component mixture split, behavior was outcome-dependent:
intensity had a beneficial but uncertain point estimate, whereas pleasantness did not
improve.

These findings define an empirical boundary, not a broad failure of molecular
representation learning. Predictive performance, perceptual alignment, incremental
information beyond chemistry, cross-dataset replication, and mixture transfer should
be evaluated separately and interpreted only for the representations, targets, and
splits examined.

\bibliographystyle{plainnat}
\bibliography{references}

\clearpage
\appendix
\section*{Supplementary Material}
\addcontentsline{toc}{section}{Supplementary Material}
\renewcommand{\thetable}{S\arabic{table}}
\setcounter{table}{0}
\renewcommand{\thefigure}{S\arabic{figure}}
\setcounter{figure}{0}
\newcommand{\code}[1]{\path{#1}}
\newcommand{\hashfour}[4]{\texttt{#1}\allowbreak{}\texttt{#2}\allowbreak{}\texttt{#3}\allowbreak{}\texttt{#4}}
\sloppy
\raggedbottom
\setlength{\emergencystretch}{2em}
\setcounter{topnumber}{4}
\setcounter{bottomnumber}{2}
\setcounter{totalnumber}{6}
\setcounter{dbltopnumber}{3}
\renewcommand{\topfraction}{0.95}
\renewcommand{\bottomfraction}{0.85}
\renewcommand{\textfraction}{0.05}
\renewcommand{\floatpagefraction}{0.80}
\renewcommand{\dbltopfraction}{0.95}
\renewcommand{\dblfloatpagefraction}{0.80}
\section{Scope and Relationship to the Main Paper}

The main text is self-contained and remains the authoritative statement of
the primary claims. The supplementary material provides additional dataset,
methodological, sensitivity, and reproducibility details.

\begin{table}[H]
\centering\small
\renewcommand{\arraystretch}{0.82}
\caption{Scientific roadmap for the supplementary analyses.}
\begin{tabular}{p{0.19\linewidth}p{0.23\linewidth}p{0.48\linewidth}}
\toprule
Main-paper claim & Scientific evidence & Evidence in this supplement \\
\midrule
Human geometry is reproducible & participant split & Table S5 and Figure S1 \\
Model--human geometry is weak & molecule/RSA & Table S6 and Figure S1 \\
Human geometry partially agrees across protocols & shared molecules & Cross-Dataset Human Geometry section \\
No clear increment beyond RDKit--Morgan & molecule/out-of-fold prediction & Table S8 and Figure S2 \\
Mixture effects depend on outcome and encoder & strict mixture unit & Tables S9--S10 and Figure S2 \\
Strict split qualification & candidate component set & Candidate Partition Analysis section \\
\bottomrule
\end{tabular}
\end{table}
\section{Dataset Acquisition and Sources}

The study uses public secondary datasets only
\cite{keller_vosshall_2016,bierling_scientific_data_2025,bierling_2025,ma_binary_mixtures_2021}.
No participant was recruited for this work.
The Bierling odor table has 74 source odor codes. \code{4Isoprop} (cuminol;
CID 325) has no rating row, leaving 73 resolved molecules in molecule-level
analyses. Missing attribute entries are handled during molecule aggregation;
the primary cohort contains 10,805 rows complete on all three attributes. The
remaining 385 rows are missing all three rating fields (each attribute therefore
has 385 missing values). Raw files are intentionally
absent from the Code and Data Supplement because redistribution permission and
source access conditions remain with the original providers.

\section{Cohort Definitions}

\textbf{Keller.} Eligibility follows the source study population: 55
participants and the final 476-molecule set with the three
target attributes. Participant means are aggregated for each molecular
identity.

The Keller configuration records participant ID, odor dilution, and vial
number. Outcome construction drops missing values separately for each attribute
and averages the available ratings within molecule and dilution; these source-design
fields were not used as predictors.

For Bierling, the primary cohort follows source-design fields rather than
model performance: \code{study == main}, \code{inclusion == 1}, sampling group
not patient, and no retest rows. It includes home and laboratory non-patient
sampling groups, 1,119 participant identifiers, 11,190 rating rows, and the
same 73 molecules. No outcome or model statistic selected the cohort.

\textbf{Bierling cohort sensitivity.} A previous broader cohort contained every mapped main/retest row
on the final 73-molecule geometry index: 1,314 participant identifiers and
13,260 rows. It is used only for the secondary analysis described in the
section entitled ``Bierling Cohort-Definition Sensitivity.''

For Ma, 30 trained assessors contribute 6,660 participant-level rows. Ratings
are aggregated into 222 stimulus-pair/replicate units spanning 198 unique
unordered component pairs. Molecular-component identities, not row positions,
determine strict train/test membership.

\begin{table}[H]
\centering\small
\setlength{\tabcolsep}{2pt}
\caption{Scientific dataset summary. Source access and redistribution
conditions remain with the original providers.}
\begin{tabular}{p{0.12\linewidth}p{0.20\linewidth}p{0.17\linewidth}p{0.22\linewidth}p{0.22\linewidth}}
\toprule
Dataset & Source and access & Participants or assessors & Molecules, components, and analysis units & Outcomes and analysis construction \\
\midrule
Keller--Vosshall & Public source conditions and cited study & 55 eligible participants & 476 molecules; molecule-level means & Participant ratings were aggregated by molecule; intensity, pleasantness, and familiarity define the three-dimensional perceptual profile. \\
Bierling & Public Zenodo/source conditions and cited data paper & 1,119 primary-cohort participants & 73 molecule analysis units and 11,190 rating rows & Intensity, pleasantness, and familiarity are aggregated by molecule under source-design cohort rules. \\
Ma & Public source conditions and cited dataset paper & 30 trained assessors; 6,660 participant-level rows & 72 components; 222 stimulus-pair/replicate units; 198 unique unordered pairs & Assessor ratings aggregate to unit-level \(I_{AB}\) intensity and \(P_{AB}\) pleasantness; raw files are not redistributed. \\
\bottomrule
\end{tabular}
\end{table}

\section{Molecular Identity Resolution and Standardization}

Source records are parsed into a molecular registry containing
isomeric canonical SMILES, non-stereo canonical SMILES, InChI, InChIKey,
connectivity key, formula, representative name/CID, and source membership.
RDKit-parsed isomeric canonical SMILES provide the canonical representation.
Exact stereo-aware registry identifiers are the primary join keys. A separate
stereo-agnostic identifier is used for identity-resolution checks only; it does not replace the
primary matching policy.

Multiple source rows resolving to one identity are mapped before aggregation.
Unresolved identities do not enter geometry, prediction, or mixture analyses.
The registry preserves the parsed charge, aromaticity, and stereochemical
representation supplied by RDKit. A distinct, general salt-stripping,
neutralization, or tautomer-normalization policy was not separately recorded;
none is inferred here. Duplicate handling and all joins use registry keys
rather than row order.

Leakage safeguards operate at molecular identity: single-molecule folds group
by registry ID; the 63 cross-dataset molecules use exact stereo-aware matches
in identical order; and strict mixture membership is assigned by component
identity, excluding every mixture spanning training and held-out components.

\begin{table}[H]
\centering\small
\caption{Identity and matching summary.}
\begin{tabular}{p{0.38\linewidth}r p{0.31\linewidth}}
\toprule
Dataset entity & Count & Status \\
\midrule
Keller analysis identities & 476 & resolved, stereo-aware \\
Bierling source odor codes & 74 & one lacks ratings \\
Bierling analysis identities & 73 & resolved, stereo-aware \\
Ma components & 72 & component-level registry joins \\
Keller--Bierling shared identities & 63 & exact stereo-aware order \\
\bottomrule
\end{tabular}
\end{table}
All 72 Ma components and all 222 mixture units were resolved. Bierling has 74
source odor codes, of which 73 have rating rows, and the final Keller analysis
uses 476 resolved identities. Identifier conflicts were not merged
automatically; exact stereo-aware identities remain separate when stereo is
explicit versus unspecified.

\section{Representation Provenance}

Both neural representations were extracted once from canonical
SMILES under deterministic evaluation settings and were never fine-tuned.
The full locally verified checkpoint revisions are given below
\cite{ross_molformer_2022,chithrananda_chemberta_2020}.

\begin{table}[H]
\centering\small
\setlength{\tabcolsep}{2pt}
\caption{Representation provenance and scientific roles.}
\resizebox{\textwidth}{!}{%
\begin{tabular}{p{0.14\linewidth}p{0.22\linewidth}p{0.22\linewidth}p{0.17\linewidth}p{0.19\linewidth}}
\toprule
Representation & Identifier and revision & Input and pooling & Dimension or features & Evaluation and scientific role \\
\midrule
MoLFormer &
\code{ibm-research/MoLFormer-XL-both-10pct};
\hashfour{a14249e5ad9e3e7c}{3b1bb604393e914c}{fcebd2c8}{} &
canonical SMILES; checkpoint-associated tokenizer (distinct revision not separately recorded); model pooled output &
768-dimensional embedding &
deterministic evaluation; no fine-tuning; CPU extraction; primary encoder for incremental prediction, geometry, and mixtures \\
ChemBERTa &
\code{DeepChem/ChemBERTa-77M-MLM};
\hashfour{ed8a5374f2024ec8}{da53760af91a33fb}{8f6a15ff}{} &
canonical SMILES; checkpoint-associated tokenizer (distinct revision not separately recorded); attention-mask-weighted mean of final hidden states &
384-dimensional embedding &
deterministic evaluation; no fine-tuning; CPU extraction; secondary encoder for geometry and mixtures \\
RDKit & RDKit 2025.09.2 & canonical molecular descriptors; cosine geometry preprocessing &
217 2D descriptors &
deterministic evaluation; geometry and chemistry baseline \\
Morgan & RDKit 2025.09.2 & canonical molecular fingerprints; radius 2, 2,048 bits, chirality enabled &
2,048 binary features &
deterministic evaluation; Tanimoto geometry and chemistry baseline \\
\bottomrule
\end{tabular}%
}
\end{table}
\FloatBarrier

The same Morgan fingerprints were used across geometry and prediction. Geometry
uses \(1-\)Tanimoto similarity; prediction leaves binary bits unscaled.
The RDKit descriptor matrix contained 217 descriptors, including 8 nonfinite
entries and 36 constant descriptors. Geometry replaces
nonfinite entries by descriptor medians, population-standardizes columns, and
sets residual constant-column nonfinite values to zero. Prediction performs
imputation and scaling inside each training fold; Morgan bits remain unscaled.
We could not determine whether encoder pretraining data overlapped with Keller,
Bierling, or Ma, so we do not claim that the datasets were absent from
pretraining.

\section{Construction of Human Perceptual Geometry}

Ratings are aggregated to molecule-level intensity, pleasantness,
and familiarity. Within each dataset, each attribute is standardized across
molecules. Euclidean distance in the resulting three-attribute space defines
human geometry. Learned embeddings use cosine distance. RDKit descriptors are
median-filled and z-scored before cosine distance. Morgan fingerprints use
Tanimoto distance.

RSA is Spearman correlation of strict upper triangles:
\[
\rho_{\rm RSA}=\operatorname{Spearman}
\left(\operatorname{vec}_{\triangle}D_{\rm rep},
\operatorname{vec}_{\triangle}D_{\rm human}\right).
\]
Global RSA measures preservation of the overall rank ordering of all
molecule-pair distances. It does not identify which odor attributes drive a
relationship and is not equivalent to outcome-specific prediction.

\section{Empirical Participant Split-Half Reproducibility}

Participants are assigned to independent halves. Molecule means and
the three-attribute Euclidean geometry are independently reconstructed in each
half on the same molecular set, and strict-upper-triangle RSA compares halves.
Bierling splits are stratified by study, odor set, sampling stratum,
inclusion, and randomization stratum. Participants were split using their
unique source participant identifiers. We used master seed 20260715 and an
independent child seed for the Bierling split procedure. One thousand valid
splits were obtained for each dataset; no primary-cohort Bierling split was
rejected.

\begin{table}[H]
\centering\small
\caption{Empirical split-half reproducibility (median and 2.5th--97.5th
percentiles over 1,000 valid splits).}
\begin{tabular}{lrr}
\toprule
Dataset & Median RSA & Interval \\
\midrule
Keller & 0.743 & [0.719, 0.768] \\
Bierling primary & 0.855 & [0.816, 0.888] \\
\bottomrule
\end{tabular}
\end{table}
These are aggregate empirical reproducibility estimates, not theoretical noise
ceilings and not model-performance ceilings.

\section{Molecule Bootstrap and Geometry Controls}

A molecule bootstrap samples molecules with replacement, reconstructs
both representation and human distance matrices, and recomputes RSA from
strict upper triangles. Each reported interval uses 2,000 valid percentile
replicates. No replicates were rejected for Bierling or the cross-dataset
analysis. Molecule pairs share endpoints and therefore are not independent
analysis units.

The geometry-null control permutes representation-matrix
molecule labels while holding the human matrix fixed and recomputes RSA.

\section{Full Single-Dataset Geometry Results}

\begin{table}[H]
\centering\small
\setlength{\tabcolsep}{3pt}
\caption{Complete model-to-human geometry results. Intervals are
95\% molecule-bootstrap percentile intervals.}
\begin{tabular}{llrrr}
\toprule
Dataset & Representation & RSA & Lower & Upper \\
\midrule
Keller & MoLFormer & 0.019266 & -0.009191 & 0.061307 \\
Keller & ChemBERTa & 0.022025 & -0.001352 & 0.059146 \\
Keller & RDKit & 0.038670 & 0.028400 & 0.063236 \\
Keller & Morgan & 0.055948 & 0.021963 & 0.104469 \\
Bierling & MoLFormer & 0.115867 & 0.050326 & 0.257747 \\
Bierling & ChemBERTa & 0.127222 & 0.064878 & 0.271521 \\
Bierling & RDKit & 0.032647 & 0.016159 & 0.145708 \\
Bierling & Morgan & 0.157641 & 0.085979 & 0.298924 \\
\bottomrule
\end{tabular}
\end{table}

Gaussian controls remained near zero; see the Negative Controls subsection.
Substantial interval overlap means that Morgan's highest
Bierling point estimate does not establish significant superiority.

\section{Bierling Cohort-Definition Sensitivity}
\label{sec:bierling-cohort-sensitivity}

This sensitivity analysis tests whether the source-design cohort
definition changes the molecule-level outcomes, human geometry, or
representation comparisons. The primary cohort was selected using source-design
criteria rather than model performance; the complete comparison is:

\begin{table}[H]
\centering\small
\caption{Bierling broader-versus-primary cohort sensitivity. Changes are
primary minus broader.}
\setlength{\tabcolsep}{2pt}
\begin{tabular}{lrrr}
\toprule
Quantity & Broader & Primary & Change/agreement \\
\midrule
Participant identifiers & 1,314 & 1,119 & -195 \\
Rating rows & 13,260 & 11,190 & -2,070 \\
Molecules & 73 & 73 & same set \\
Intensity outcome & -- & -- & Spearman 0.981 \\
Pleasantness outcome & -- & -- & Spearman 0.994 \\
Familiarity outcome & -- & -- & Spearman 0.979 \\
Human distance matrix & -- & -- & RSA 0.963 \\
Split-half median & 0.867 & 0.855 & -0.012 \\
MoLFormer RSA & 0.111 & 0.116 & +0.005 \\
ChemBERTa RSA & 0.123 & 0.127 & +0.004 \\
RDKit RSA & 0.029 & 0.033 & +0.003 \\
Morgan RSA & 0.151 & 0.158 & +0.007 \\
Shared-geometry RSA & 0.352 & 0.331 & -0.021 \\
\bottomrule
\end{tabular}
\end{table}
The outcomes and human geometry remained highly concordant across cohort
definitions, and the representation-level conclusions were unchanged. The
primary cohort was defined using study-design variables rather than model
performance.

\section{Cross-Dataset Human Geometry}

Sixty-three exact shared stereo-aware molecules are placed in
identical order. Keller and Bierling attributes are standardized independently
within dataset. The 1,953 pairwise distances are descriptive, not independent
units. A shared-molecule bootstrap resamples 63 identities, reconstructs both
geometries, and uses 2,000 valid replicates with zero rejections.

The exact RSA is 0.330841323 with 95\% interval [0.203700, 0.507320],
reported as 0.331 [0.204, 0.507] in the main paper. This is positive but
incomplete protocol-level agreement. It is neither participant reliability nor
a ceiling for model performance.

\begin{figure}[H]
\centering
\includegraphics[width=0.96\textwidth]{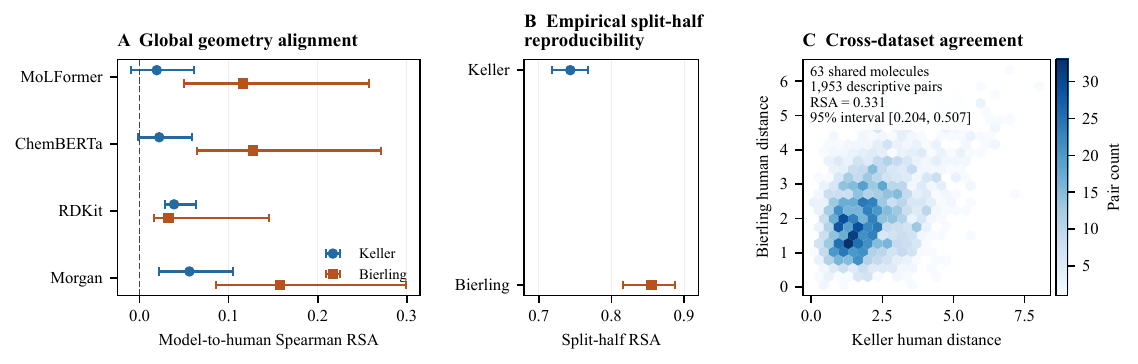}
\caption{Geometry, empirical split-half reproducibility, and cross-dataset
agreement. Reproduced from main-paper Figure 2 for reference. Molecules are the
analysis and bootstrap units for model--human and cross-dataset geometry;
participants are split for empirical reproducibility. Intervals are the
molecule-bootstrap or participant-split distributions described in the text.}
\label{fig:geometry-results}
\end{figure}

\section{Incremental Prediction Protocol}

The primary contrast is RDKit+Morgan versus
RDKit+Morgan+MoLFormer. Five molecule-level folds, ten repeats, and master seed
20260713 yield ten out-of-fold predictions per molecule. Predictions are
averaged by molecule before MAE. Training-fold preprocessing uses mean
imputation for every block, standardization for RDKit and MoLFormer, and
unscaled binary Morgan bits. Blocks are concatenated without weighting.
Ridge \(\alpha=10\) is fixed by the existing analysis policy.

The incomplete-baseline sensitivity compares RDKit with
RDKit+MoLFormer. The nonlinear sensitivity uses 80-tree Random Forests,
maximum depth 10, minimum leaf size 2, \code{max_features=sqrt}, and seed
20260713. Paired uncertainty resamples molecules from averaged out-of-fold
predictions 2,000 times. The sign convention is
\[
\Delta{\rm MAE}={\rm MAE}(\text{combined})-{\rm MAE}(\text{chemistry}),
\]
so negative values are beneficial.

\section{Complete Incremental Prediction Results}

\begin{table}[H]
\centering\small
\caption{Complete incremental prediction results. Intervals are
paired molecule-bootstrap intervals.}
\setlength{\tabcolsep}{3pt}
\resizebox{\textwidth}{!}{%
\begin{tabular}{p{0.14\linewidth}p{0.08\linewidth}p{0.22\linewidth}rrrr}
\toprule
Role & Dataset & Contrast & Baseline MAE & Combined MAE & \(\Delta\)MAE & 95\% interval \\
\midrule
Primary Ridge & Keller & RDKit+Morgan \(+\) MoLFormer & 13.3687 & 13.3677 & -0.0010 & [-1.9112, 1.3650] \\
Primary Ridge & Bierling & RDKit+Morgan \(+\) MoLFormer & 9.7340 & 10.1106 & +0.3766 & [-3.2401, 2.7838] \\
Nonlinear sensitivity & Keller & RDKit+Morgan \(+\) MoLFormer & 11.3662 & 11.5664 & +0.2002 & [-0.0280, 0.4319] \\
Nonlinear sensitivity & Bierling & RDKit+Morgan \(+\) MoLFormer & 8.3799 & 8.9556 & +0.5757 & [0.0466, 1.0835] \\
RDKit-only sensitivity & Keller & RDKit \(+\) MoLFormer & 14.1474 & 13.9847 & -0.1628 & [-2.2823, 1.3661] \\
RDKit-only sensitivity & Bierling & RDKit \(+\) MoLFormer & 11.1867 & 10.2851 & -0.9016 & [-7.0170, 2.8855] \\
\bottomrule
\end{tabular}%
}
\end{table}
The Random Forest sensitivity estimates were non-beneficial: adding MoLFormer
increased MAE in both datasets, with the Bierling interval excluding zero
under this specific model, outcome, cohort, and analysis specification. In
contrast, the RDKit-only Ridge point estimates were directionally beneficial,
but both intervals crossed zero and therefore did not establish improvement.
The RDKit-only comparison is secondary because it omits the Morgan fingerprint
baseline. The primary RDKit+Morgan Ridge intervals also cross zero.

\section{Strict Mixture Split Construction}

The Ma graph has 72 components. The prespecified strict split assigns
52 components to training and 11 to held-out status with zero overlap. It
contains 101 training units and 20 test units; 101 mixtures spanning both sides
are excluded. Component membership is checked through fixed registry IDs.
Component vectors are mean-pooled under the primary composition rule. Separate
Ridge analyses target \(I_{AB}\) intensity and \(P_{AB}\) pleasantness. The
evidence is limited to this one prespecified split.

\begin{table}[H]
\centering\small
\caption{Strict-split summary.}
\begin{tabular}{lr}
\toprule
Quantity & Count \\
\midrule
Total components & 72 \\
Training components & 52 \\
Held-out components & 11 \\
Component overlap & 0 \\
Training units & 101 \\
Test units & 20 \\
Excluded cross-partition units & 101 \\
\bottomrule
\end{tabular}
\end{table}
\FloatBarrier
\section{Candidate Partition Analysis}

No second qualifying partition was found in the fixed outcome-blind pool of
5,000 candidate assignments. Among 4,936 unique test-component sets, 66 had
exactly 11 held-out components, but none reached 20 strict test units; the
maximum was 14. A qualifying partition required exactly 11 held-out
components, at least 20 strict test units, at least 101 training units, zero
component overlap, exclusion of cross-partition mixtures, and a unique
test-component set. We neither relaxed these thresholds nor fit a second
model. This search was limited to the fixed candidate pool and does not
establish nonexistence in the complete combinatorial space.

\section{Complete Mixture Results}

\begin{table}[H]
\centering\small
\caption{Complete strict-split mixture results. Standalone encoder MAEs are
included for context; \(\Delta\)MAE always compares RDKit with the combined
model.}
\resizebox{\textwidth}{!}{%
\begin{tabular}{llrrrrr}
\toprule
Outcome & Encoder & RDKit MAE & Encoder alone & RDKit+encoder & \(\Delta\)MAE & 95\% interval \\
\midrule
\(I_{AB}\) & MoLFormer & 0.4395 & 0.3157 & 0.3226 & -0.1169 & [-0.2347, 0.0118] \\
\(I_{AB}\) & ChemBERTa & 0.4395 & 0.3409 & 0.4050 & -0.0344 & [-0.1139, 0.0478] \\
\(P_{AB}\) & MoLFormer & 0.5998 & 0.5710 & 0.6248 & +0.0251 & [-0.2293, 0.2769] \\
\(P_{AB}\) & ChemBERTa & 0.5998 & 0.6518 & 0.4761 & -0.1237 & [-0.2455, 0.0085] \\
\bottomrule
\end{tabular}%
}
\end{table}

Table S10 compares RDKit alone, each encoder alone, and RDKit plus encoder.
Standalone performance does not imply complementary information; the combined
model is the relevant comparison for incremental contribution. Every
combined-model interval crosses zero, so evidence is outcome- and
representation-dependent, limited to one strict split, and establishes no
robust encoder ranking.

\FloatBarrier
\clearpage
\begin{figure}[H]
\centering
\includegraphics[width=0.96\textwidth]{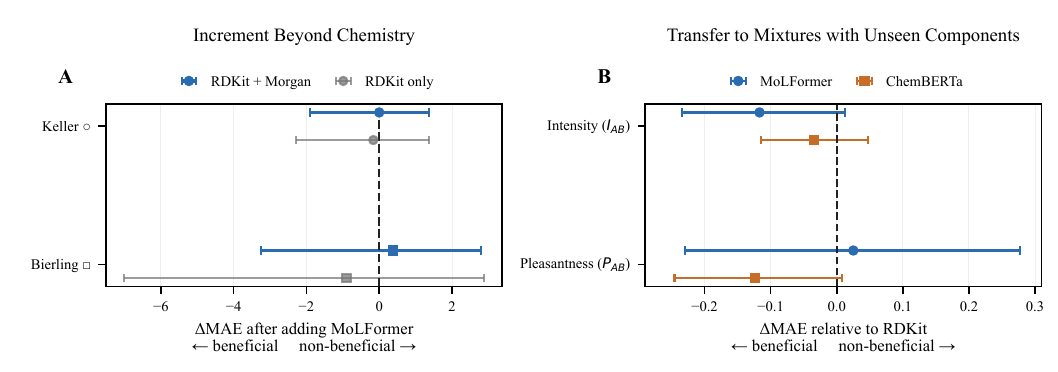}
\caption{Incremental prediction and strict-mixture transfer. Reproduced from
main-paper Figure 3 for reference. Panel A uses molecule-level averaged
out-of-fold predictions and molecule-bootstrap intervals for the primary
RDKit+Morgan Ridge comparison. Panel B uses mixture units and mixture-unit
bootstrap intervals for one prespecified strict unseen-component split.
Negative \(\Delta\)MAE is beneficial.}
\label{fig:prediction-mixture}
\end{figure}
\subsection{Sensitivity Synthesis}

The conclusions are stable to the Bierling cohort definition: the molecule set
is unchanged, outcomes and geometry are highly concordant, and the qualitative
interpretation is unchanged. Apparent MoLFormer value depends on baseline
completeness; the tested nonlinear probe did not recover incremental value, and
strict-mixture complementarity remains uncertain under the one split.

\subsection{Negative Controls}

Gaussian embedding controls remained near zero (\(-0.011802\),
\(-0.004506\), and \(-0.031643\)). The near-zero controls are consistent with
high embedding dimensionality alone not producing the observed positive RSA
values; they do not definitively rule out every dimensionality or distance-
matrix artifact.

\section{Excluded Analysis}

We report only analyses that passed the stated validation checks. An
exploratory predictive permutation procedure was excluded because its null
construction was invalid; none of its numerical outputs or p-values is used.
Morgan geometry was computed using the validated numeric-bit Tanimoto
implementation.

\section{Reproducibility}

The core analyses were run with Python 3.9.6 and RDKit 2025.09.2. The
reliability analysis additionally recorded NumPy 1.26.4 and pandas 2.3.3. The
accompanying package provides a portable Python 3.12 verification environment,
with RDKit 2025.09.2 and the required scientific Python packages. Because raw
datasets and pretrained model weights are not redistributed, package-level
verification covers the included derived artifacts rather than full
end-to-end reproduction. The Code and Data Supplement includes result tables,
configurations, mappings, split assignments, software specifications, and
package-level verification procedures.

\section{Extended Limitations}

All datasets are public secondary resources with inherited consent, sampling,
population, aggregation, and governance limitations. Cohort definitions differ
across protocols. Human geometry uses only intensity, pleasantness, and
familiarity and is not a complete olfactory representation. Rating scales,
dilution, concentration, participant design, and molecule composition vary,
limiting direct and causal interpretation.

Only generic molecular encoders were evaluated. We did not evaluate
olfaction-specific, receptor-informed, graph-based, or three-dimensional
representations. Mixture composition uses mean pooling and omits
concentration and interaction-aware mechanisms. The single strict split has
20 test units, which remain dependent through shared components within the test
set. No second qualifying partition was found in the fixed outcome-blind pool
of 5,000 candidate assignments; this does not establish nonexistence in the
complete combinatorial space.

Global RSA summarizes all pairwise rank orderings and does not localize useful
outcome information. Low RSA does not prove absence of predictive signal.
Incremental validity depends on baseline completeness, preprocessing, probe,
split, and outcome; apparent gains against RDKit alone can disappear against
RDKit+Morgan. These analyses do not support causal claims or a universal
ranking of molecular representations.

\end{document}